# Evidence for metastable photo-induced superconductivity in K$_3$C$_{60}$

M. Budden[1], T. Gebert[1], M. Buzzi[1], G. Jotzu[1], E. Wang[1], T. Matsuyama[1],
G. Meier[1], Y. Laplace[1], D. Pontiroli[2], M. Riccò[2], F. Schlawin[3], D. Jaksch[3], A. Cavalleri[1,3,*]

[1] Max Planck Institute for the Structure and Dynamics of Matter, 22761 Hamburg, Germany
[2] Dipartimento di Scienze Matematiche, Fisiche e Informatiche, Università degli Studi di Parma, Italy
[3] Department of Physics, Clarendon Laboratory, University of Oxford, Oxford OX1 3PU, United Kingdom
* e-mail: andrea.cavalleri@mpsd.mpg.de

**Far and mid infrared optical pulses have been shown to induce non-equilibrium unconventional orders in complex materials, including photo-induced ferroelectricity in quantum paraelectrics[1,2], magnetic polarization in antiferromagnets[3] and transient superconducting correlations in the normal state of cuprates and organic conductors[4-9]. In the case of non-equilibrium superconductivity, femtosecond drives have generally resulted in electronic properties that disappear immediately after excitation, evidencing a state that lacks intrinsic rigidity. Here, we make use of a new optical device to drive metallic K$_3$C$_{60}$ with mid-infrared pulses of tunable duration, ranging between one picosecond and one nanosecond. The same superconducting-like optical properties observed over short time windows for femtosecond excitation are shown here to become metastable under sustained optical driving, with lifetimes in excess of ten nanoseconds. Direct electrical probing becomes possible at these timescales, yielding a vanishingly small resistance. Such a colossal positive photo-conductivity is highly unusual for a metal and, when taken together with the transient optical conductivities, it is rather suggestive of metastable light-induced superconductivity.**

Alkali-doped fullerides of the $A_3C_{60}$ family (Figure 1a) exhibit high temperature superconductivity[10-12], which is generally tuned with chemical or physical pressure[13-15] (see Figure 1b). Recent work[6,7] has focused on the dynamical manipulation of these materials with optical pulses at mid-infrared frequencies, which were tuned close to resonance with local vibrations of the $C_{60}$ molecules to induce superconducting correlations above the thermodynamic transition temperature.

These experimental results[6,7] are summarized in Figure 1c and 1d for a new batch of fulleride samples. $K_3C_{60}$ powders were held at a base temperature of $T = 100\,K \gg T_c = 20\,K$ and irradiated with 100 fs-long, 7.3 μm-wavelength ($\hbar\omega \sim 170$ meV) pulses at a fluence of 3 mJ/cm², yielding a short-lived transient state with large changes in the terahertz optical properties. The transient optical response was probed with phase-sensitive time-domain terahertz spectroscopy, which directly yields the real and imaginary part of the reflectance and hence can be used to retrieve the complex optical conductivity without the need of Kramers-Kronig transformations (see Supplementary Information S4 and Refs. 6,7). The transient low-frequency optical conductivity induced by femtosecond excitation was almost indistinguishable from that of the equilibrium superconducting state measured in the same material at $T \ll T_c = 20\,K$ (cf. Figure 1c). Such "superconducting-like" optical properties consist of a perfect low energy reflectance (R = 1), a vanishingly small real part of the optical conductivity $\sigma_1(\omega)$ for all photon energies lower than the energy gap 2Δ and an imaginary conductivity $\sigma_2(\omega)$ that diverges toward low frequencies as $\sim 1/\omega$, which is itself indicative of a large zero frequency conductivity.

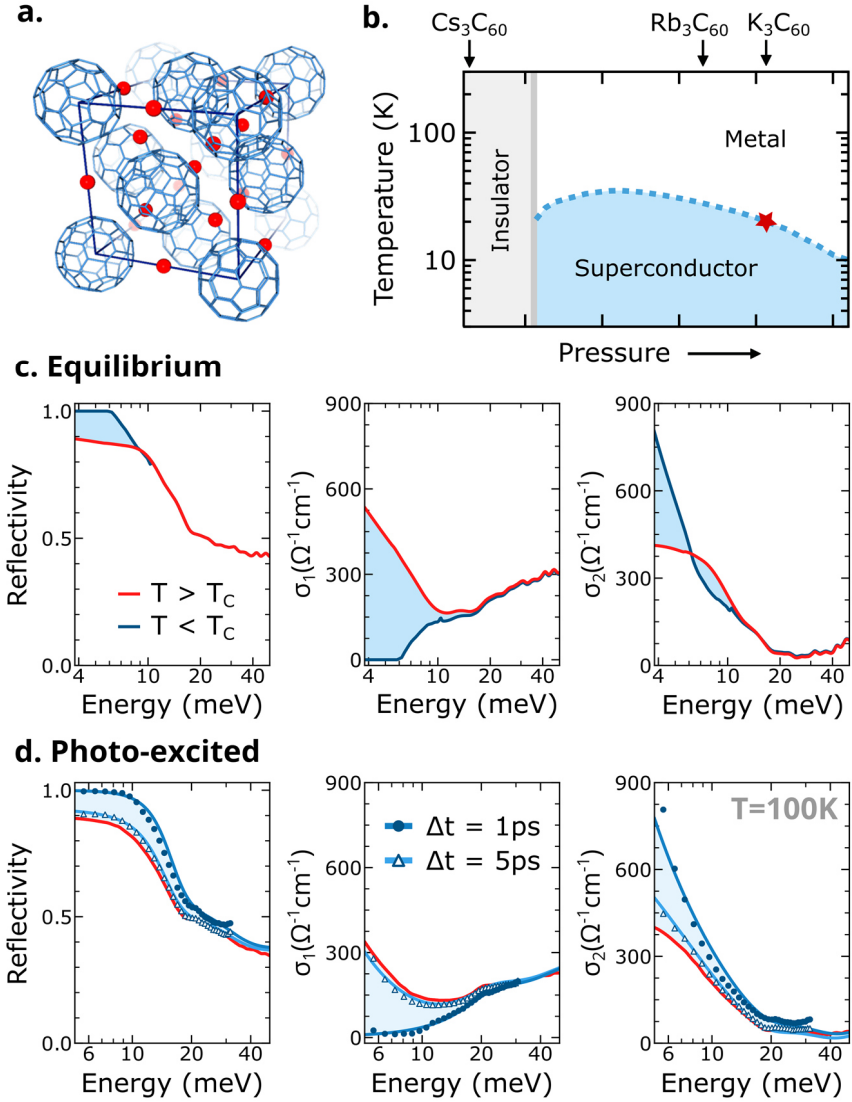

**Figure 1: Crystal structure, phase diagram, equilibrium phase transition, and short-lived light induced phase in $K_3C_{60}$. a.** Crystal structure of the molecular superconductor $K_3C_{60}$. $C_{60}$ molecules are arranged in a face centered cubic lattice. Potassium atoms (red) occupy the interstitial voids. **b.** Phase diagram of the $A_3C_{60}$ alkali doped fullerene family. The ground state of the material can be tuned by chemical or physical pressure. The star indicates the $K_3C_60$ compound investigated in the present work becomes superconducting for $T < T_c = 20\ K$. **c.** Equilibrium reflectivity (sample-diamond interface), real, and imaginary part of the optical conductivity of $K_3C_{60}$ measured upon cooling across the equilibrium superconducting transition. **d.** Same quantities measured at equilibrium (red lines), 1 ps (blue filled dots), and 5 ps (light blue triangles) after photoexcitation. The light and dark blue lines are Drude-Lorentz fit to the transient optical data (see Supplementary Information for more details). These data were acquired at a base temperature $T = 100\ K$ with a fluence of 3 mJ/cm$^2$.

These observations have generated interest as they may open up the possibility of achieving photo-induced superconducting states[4-8,16,17] at or in the vicinity of room temperature.

However, all the experiments reported to date have evidenced states that disappear immediately after optical excitation (see optical properties measured at 5 ps time delay in Figure 1d). These short lifetimes would prevent most applications, and have even raised controversy on the interpretation of the data itself[18,19].

Here, we explore the possibility of longer-lived superconductivity under a sustained optical drive. We first modified the experimental setup used for the experiments shown in Figure 1 and lengthened the 7.3 µm-wavelength pump pulses by making them propagate in a dispersive $CaF_2$ rod (see Figure 2a for a schematic of the experimental setup). These chirped pump pulses had a duration of ~1 ps, enabling a sixfold increase in the pulse energy (up to 18 mJ/cm²). Because the pulse duration was made longer, these experiments could be conducted at pulse energies that would have damaged the sample for femtosecond pulse durations, and hence enabled a new regime of excitation.

The pump-induced changes in the low frequency reflectivity and complex optical conductivity, measured on the same sample and at the same temperature $T = 100$ K of the experiments of Figure 1, are displayed in Figure 2b. Representative reflectivity spectra $R(\omega)$ and complex optical conductivity $\sigma_1(\omega) + i\sigma_2(\omega)$ measured for a 1-ps pump-pulse duration and 18 mJ/cm² fluence are reported for time delays of –5 ps (red curve), 10 ps, 300 ps, and 12 ns (blue curves in Figure 2b). These plots evidence a similar response as that reported in Figure 1d, although with a far longer relaxation time.

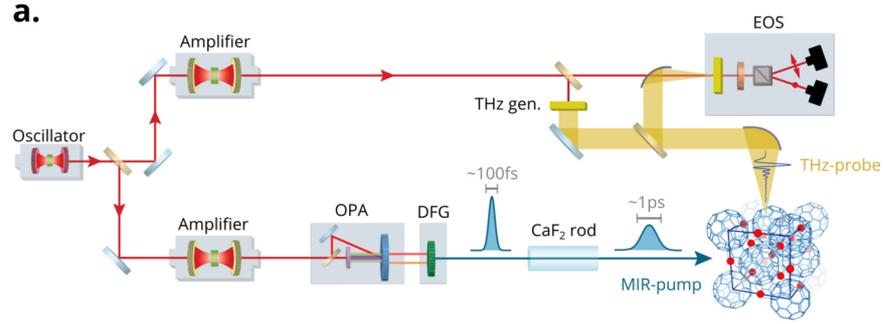

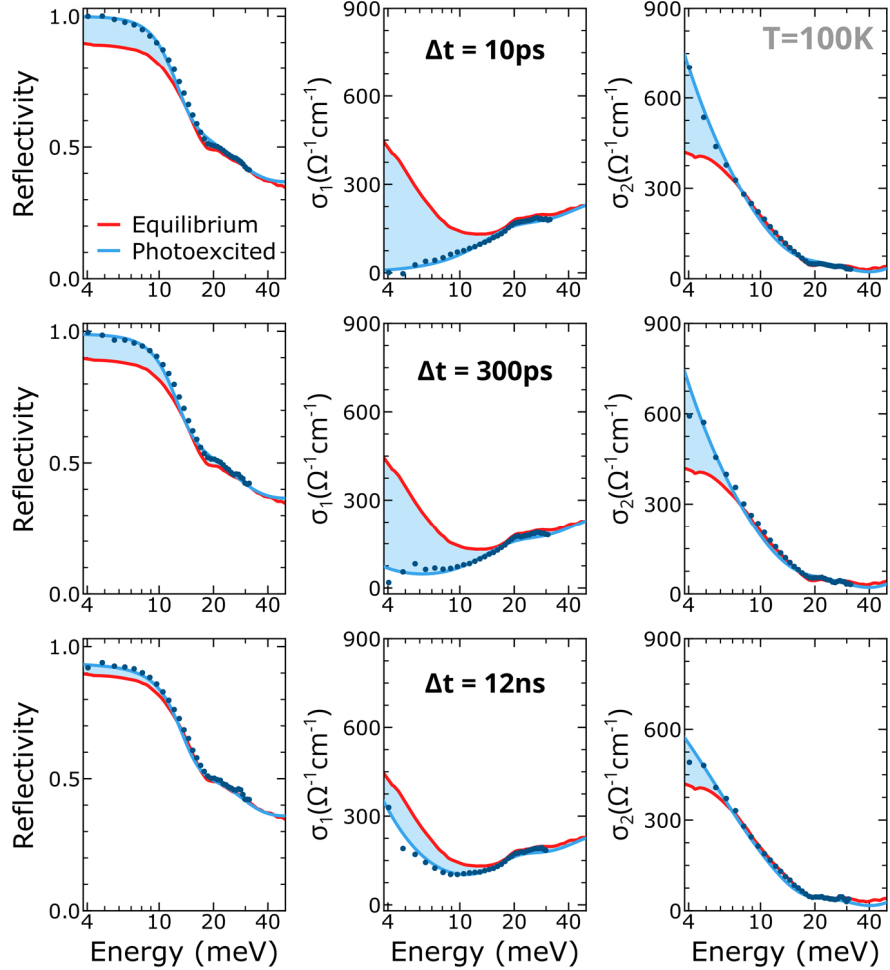

**Figure 2: Long-lived light induced phase in $K_3C_{60}$ generated with intense, 1 ps long excitation pulses**. **a.** Schematic of the experimental setup for pump-probe experiments producing 1 ps long pump pulses centered at 7.3 μm wavelength. The 7.3 μm wavelength, 100 fs long pulses are stretched by linear propagation in a highly dispersive $CaF_2$ rod. The photoinduced changes in the optical response of $K_3C_{60}$ upon irradiation are detected with transient THz time domain spectroscopy. **b.** Equilibrium reflectivity (sample-diamond interface), real, and imaginary part of the optical conductivity of $K_3C_{60}$ measured at equilibrium (red lines), 10 ps, 300 ps, and 12 ns (blue filled symbols) after photoexcitation. The dark blue lines are Drude-Lorentz fits to the transient optical data (see Supplementary Information for more details). These data were acquired at a base temperature $T = 100\ K$ with a fluence of 18 mJ/cm² and a pump pulse duration of 1 ps.

All features, which characterize the light-induced superconducting-like state, that is perfect reflectivity ($R = 1$), gapped $\sigma_1(\omega)$ and a divergent $\sigma_2(\omega)$, were observed for these longer pulses with higher integrated excitation intensity to persist up to at least 300 ps after excitation.

A more comprehensive exploration of sustained optical driving, beyond the limited pulse-width tuneability of the setup discussed in Figure 2, is reported in Figures 3, 4, and 5. These measurements were made possible by the development of a new optical device, based on a $CO_2$-gas laser that was optically synchronized to a femtosecond $Ti:Al_2O_3$ laser, and that delivered 10.6 μm wavelength pulses with durations that could be tuned between ~5 ps and ~1 ns (see Figure 3a).

Near infrared femtosecond pulses from a $Ti:Al_2O_3$ laser were converted to 10.6 μm wavelength in an optical parametric amplifier, and used to seed a $CO_2$ laser oscillator synchronizing the two lasers in time. The oscillator emitted trains of nanosecond-long pulses out of which the most intense was selected by a Pockels cell and amplified to 10 mJ energy in a second, multi-pass $CO_2$ laser amplifier. The duration of these amplified pulses was then tuned as displayed in Figure 3a. The "front" and the "back" of the 1-ns-long $CO_2$ pulses were "sliced" using a pair of photo-excited semiconductor wafers as plasma mirrors. Because the wafers were set at Brewster's angle and their bandgap was much larger than the ~120 meV photon energy of the $CO_2$ laser, they were almost perfectly transparent when unexcited. As two femtosecond optical pulses struck each one of the wafers at adjustable time delays, these became reflective due to the injection of dense electron hole plasmas. The wafer pair then transmitted 10.6 μm pump pulses with a duration tunable from ~5 ps to ~1 ns (see Supplementary Information S3). These pulses were then used to pump the $K_3C_{60}$ sample, which was probed with the same time-domain terahertz reflectivity probe used for the experiments of Figures 1 and 2.

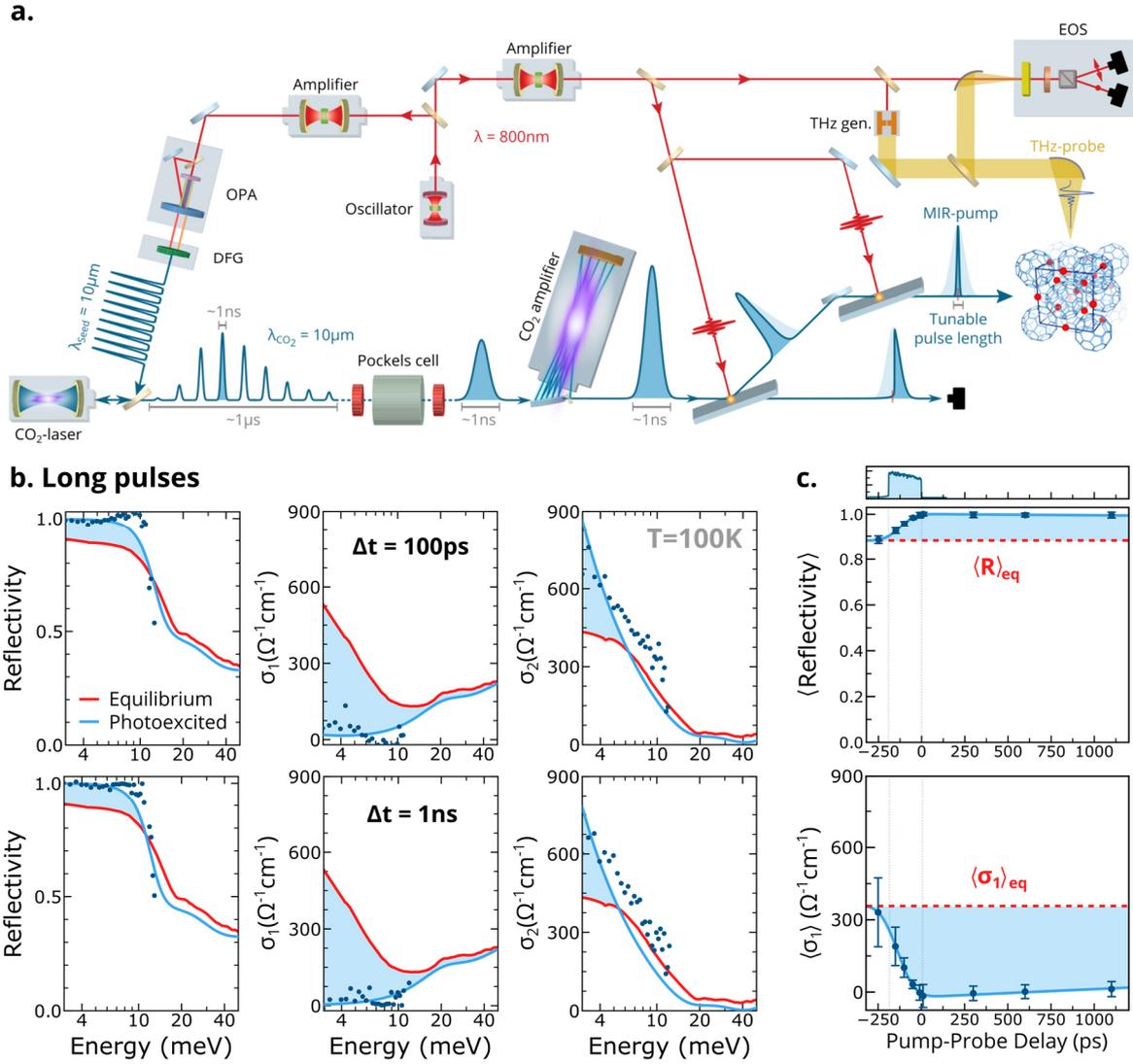

**Figure 3: Long-lived light induced phase in $K_3C_{60}$ generated with intense, 300 ps long excitation pulses**. **a.** Schematic of the experimental setup for pump-probe experiments producing picosecond pulses of duration variable between 5 ps and 1.3 ns, centered at 10.6 μm wavelength, and with a pulse energy of up to 10 mJ. The pulses are generated by seeding a $CO_2$ gas laser with a femtosecond source. After amplification in a second $CO_2$ gas laser, their duration is tuned by semiconductor slicing. The photoinduced changes in the optical response of $K_3C_{60}$ upon irradiation are detected with transient THz time domain spectroscopy. **b.** Equilibrium reflectivity (sample-diamond interface), real, and imaginary part of the optical conductivity of $K_3C_{60}$ measured at equilibrium (red lines), 100 ps, and 1 ns (blue filled symbols) after photoexcitation. The dark blue lines are Drude-Lorentz fits to the transient optical data (see Supplementary Information for more details). These data were acquired at a base temperature $T = 100\ K$ with a fluence of 53 mJ/cm² and a pump pulse duration of 300 ps. **c.** Time dependence of the average reflectivity and spectral weight in the real part of the optical conductivity $\sigma_1(\omega)$ in the region of the photoinduced gap (2-10 meV). These data were acquired at a base temperature $T = 100\ K$ with a fluence of 25 mJ/cm² and a pump pulse duration of 200ps. The top panel shows the measured time profile of the excitation pulse.

Figure 3b displays snapshots of the transient optical properties ($R(\omega), \sigma_1(\omega), \sigma_2(\omega)$) measured for $K_3C_{60}$ at T = 100 K, before (red curves), 100 ps, and 1 ns after photoexcitation (blue symbols, top and bottom panels respectively) with a 300 ps long pulse centered at 10.6 μm wavelength. Because the repetition rate of these measurements was 18 Hz, the signal to noise was reduced compared to the cases reported in Figures 1 and 2, which were measured at 1 kHz.

Crucially, the transient optical spectra measured in these conditions showed the same superconducting-like features as reported in Figures 1d and 2b for all pump pulse durations at up to 1 ns after excitation. Note that the 10.6 μm wavelength radiated by the $CO_2$ laser is different from the 7.3 μm wavelength used in the experiments of Figures 1 and 2. However, excitation with femtosecond optical pulses at this wavelength had previously been shown to induce the same transient optical signatures generated with 7.3 μm wavelength excitation, although with a lower efficiency[6].

The time evolution of the superconducting-like terahertz optical properties is displayed in Figure 3c. The top panel reports the average reflectivity $\langle R(\omega) \rangle$ in the region where $\sigma_1(\omega)$ exhibited a gap (2-10 meV). The lower panel displays the average value of the corresponding real part of the optical conductivity $\langle \sigma_1(\omega) \rangle$, which reached zero after optical excitation, reflecting full gapping. Both quantities remained unchanged after excitation for all time delays measured up to 1 ns. Extended measurements indicate a lifetime of the light-induced superconducting state of ~10 ns (see Supplementary Information S9). From the optical spectra, we extract also an estimate of the "zero-frequency resistivity" $\rho_0 = 1 / \lim_{\omega \to 0} \sigma_1(\omega)$, which is based on a Drude-Lorentz fit to the transient optical properties (see Supplementary Information S5). This fitting procedure yielded a vanishingly small $\rho_0(t)$ for all time delays after excitation (see Figure 4a).

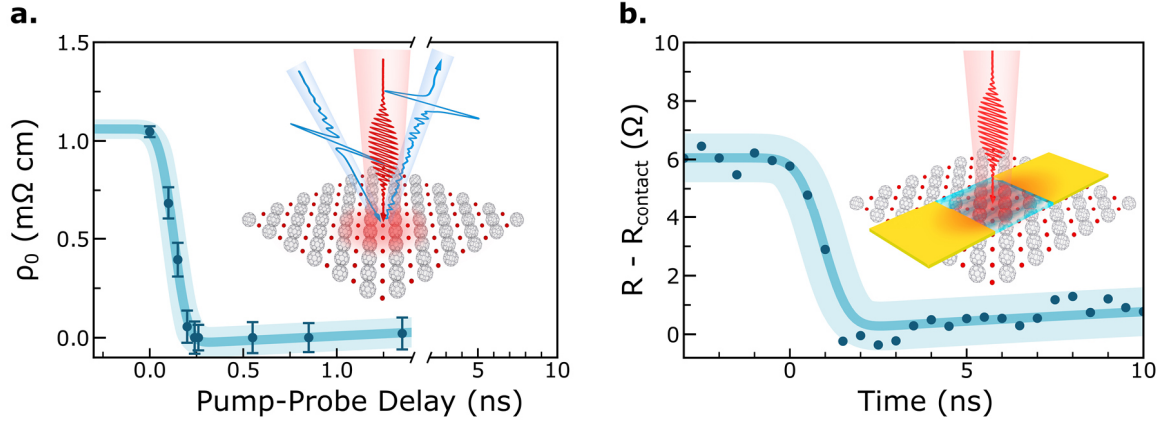

**Figure 4: Transient resistivity measurements of the long-lived light-induced phase in $K_3C_{60}$. a.** Time evolution of the transient resistivity $\rho_0$ as obtained from an extrapolation to zero frequency of a Drude-Lorentz fit to the transient optical conductivities (see Supplementary Information S5). **b.** Resistance of a laser irradiated $K_3C_{60}$ pellet embedded in a microstrip transmission line. The resistance value is obtained from a transient 2-point transport measurement where contributions from contact resistances are calibrated from a static 4-point measurement. The shaded area indicates an estimate of the systematic error introduced by this calibration (see main text and Supplementary Information S7 for more details). The insets display a schematic of the experimental geometry. These data were acquired at a base temperature $T = 100\ K$ with a fluence of $\approx 25\ mJ/cm^2$. The pump pulse durations were 200 ps and 75 ps for the optical and transport measurements, respectively.

Estimates of $\rho_0$ from optical measurements were complemented with direct electrical measurements. The $K_3C_{60}$ pellets were incorporated into lithographically patterned microstrip transmission lines. Their resistance was tracked at different times after excitation by transmitting a 1 ns voltage probe pulse that yielded time-resolved two-terminal resistance measurements (see Supplementary Information S6 and S7). The contributions due to contact resistance were normalized by performing equilibrium four-terminal measurements which were subtracted from the time-resolved resistance measurements (see Supplementary Information S7). Figure 4b shows the time evolution of the two-terminal resistance of a $K_3C_{60}$ pellet measured at T = 100 K upon photoexcitation in similar conditions to those of the optical experiments reported in Figure 3. As seen in the resistivity extrapolated from the optical measurements (cf. Figure 4a), upon excitation the resistance drops to a value that is compatible with zero and recovers on the same time scale of tens of nanoseconds extracted from the fitted results of Figure 4a (see Supplementary Information S9).

The electrical probe experiments were repeated by varying the excitation pulse duration and fluence. Figure 5a shows exemplary pump-pulse duration dependences of the measured sample photo-resistivity for excitation fluences of 1.5 mJ/cm², 10 mJ/cm², and 25 mJ/cm². The photoresistivity was mostly independent of the pump pulse duration and depended only on the total energy of the excitation pulse. This is also underscored by the data presented in Figure 5b, which illustrates the dependence of the sample resistance on the excitation fluence at constant pulse duration. Metastable zero resistance was observed for all excitation fluences in excess of 20 mJ/cm².

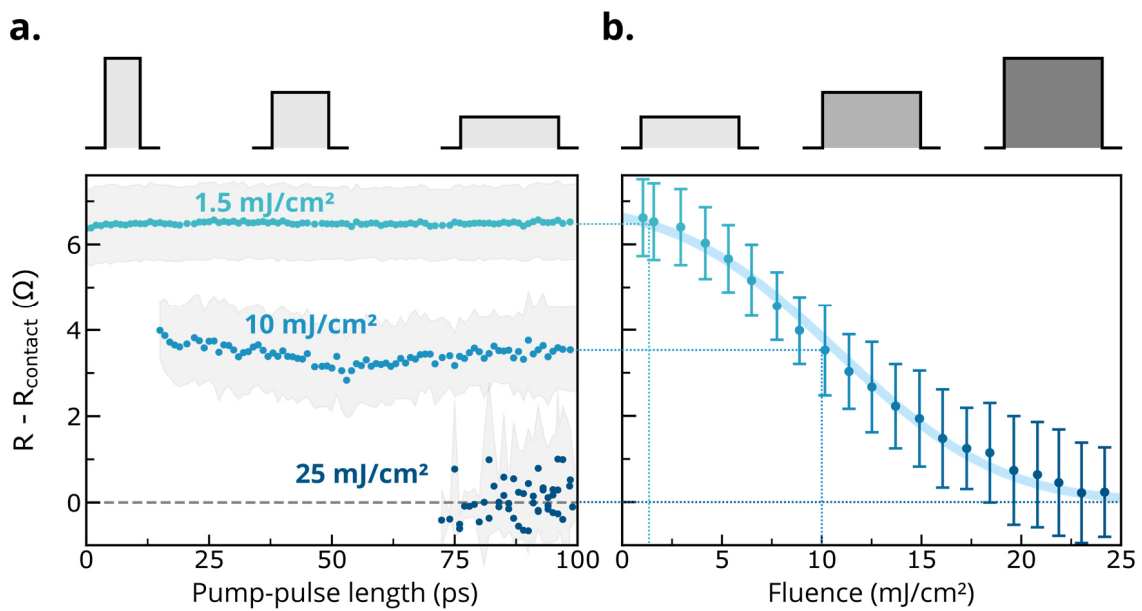

**Figure 5: Pump pulse duration and fluence dependence of the long-lived light-induced phase in $K_3C_{60}$. a.** Pulse length dependence of the resistance of a laser irradiated $K_3C_{60}$ pellet measured at three different representative fluences of 1.5 mJ/cm², 10 mJ/cm², and 25 mJ/cm². For the higher fluence values of 10 mJ/cm² and 25 mJ/cm² the minimum achievable pulse length is higher due to limitations of our experimental setup. The shading represents the standard deviation of the measurements. The sample resistance was obtained as in Figure 4 by calibrating the contributions of contact resistance in the transient 2-terminal measurement from a static 4-point measurement (see Supplementary Information S7). **b.** Corresponding fluence dependence of the calibrated sample resistance measured at a constant pump pulse duration of 95 ps. The solid line is a guide to the eye. The error bars display the standard deviation of the mean. These data were measured at a base temperature $T = 100\ K$.

These data provide evidence for a metastable state of metallic $K_3C_{60}$ with a very large positive photoconductivity. Note that even if one were to disregard the gapped terahertz conductivities and the divergent imaginary conductivity of Figure 3 as an indication of a superconducting state, the observation of such a large and positive photoconductivity would be highly unconventional for a metal, which generally exhibits photoconductivities of less than 1 %[20-22], which are often negative[23] especially when excitation is performed in the mid-infrared and far away from interband transitions. Rather, the combined observation of a vanishingly small resistance with the transient terahertz optical spectra reported in Figure 3 substantiates the assignment of a metastable superconducting state.

The discovery of a metastable state prompts comparison with previously measured responses observed under sustained driving, such as the microwave enhancement of conventional superconductivity[24,25]. Aside from the fact that the temperature scale observed here is far larger than that reported in the microwave enhancement case, we also note other key differences. Most importantly, in the case of microwave enhanced BCS superconductivity the effect was observed only for excitation below the superconducting gap, which was interpreted as a result of quasi-particle redistribution and of renormalization of the parameters entering the BCS equations[26]. In the microwave experiments, complementary measurements, where excitation was tuned immediately above the gap, yielded a reduction of the superconducting order. Here, our experiments are conducted in a different regime, that is at pump photon energies that are at least one order of magnitude larger than the low temperature equilibrium superconducting gap, and in a regime in which the primary coupling of the mid-infrared radiation is not with the condensate but with other high-energy excitations, such as molecular vibrations or collective electronic modes. Hence, it is unlikely that the mechanism invoked for microwave enhancement can explain these observations.

Other experimental reports have documented a sustained, metastable enhancement of superconductivity in oxygen deficient $YBa_2Cu_3O_{7-\delta}$ samples, after exposure to radiation with frequencies ranging from ultraviolet to near infrared[27-29]. However, in all these cases the superconducting transition temperature of the irradiated superconductor never exceeded the equilibrium transition temperature at optimal doping. These observations were either interpreted as a result of photodoping towards a more metallic state favoring superconductivity or involved annealing of oxygen deficient samples. In contrast, the data reported here show enhanced superconducting properties well above the equilibrium transition temperature by excitation with photon energies away from interband transitions. Thus, this mechanism does not seem to offer a plausible explanation for our findings, in which the response is observed in a metal and the signatures of metastable photo-induced superconductivity extend far above the equilibrium transition temperature $T_c$.

In the experiments reported here, the pulse-duration and fluence dependences of Figure 5 provide important perspective on the microscopic physics. From the observation that the amplitude of the effect is dependent only on the integrated pulse area, it becomes clear that at least in the long pulse regime the previously proposed nonlinear phonon mechanism[6], which was suggested to generate a displaced crystal structure, is most likely not correct. If this were the case, one would expect a response that depends on some power of the electric field, rather than only on the total energy of the pulse. On the other hand, mechanisms based on the parametric coupling of the light field to the electronic properties, either to amplify a superconducting order parameter[30-32] or to cool selected degrees of freedom[33,34], are not necessarily inconsistent with the observations reported here.

We also note that for the first time the photo-induced high temperature state survives far longer than the drive pulse, and hence exhibits intrinsic rigidity at timescales when the coherence is no longer being supplied by the external drive. Such rigidity can arise from microscopic correlations, and is reminiscent of the first order relaxation kinetics observed in other photo-induced phases of complex solids[1,35-38]. One can speculate that at these temperatures, relaxation of the light induced state into the true ground state of the system may proceed by nucleation of vortex anti-vortex pairs.

Significant questions remain about dissipation during the drive. Indeed, energies of 20 mJ/cm$^2$ would be expected to raise the overall temperature of the sample by at least 100 K, although hot carrier diffusion at early times[39,40] may reduce the heating to only a few tens of Kelvin. Regardless of the mechanism and the degree of heating, pre-thermalization of the driven superconducting order parameter, which may be immune to dissipation at early times, could be important.

Finally, the data reported here hold a significant promise in the quest to extend lifetimes even further. New lasers capable of generating longer pulses or suitably designed trains of pulses could be developed in order to sustain the coherence of this state. Extended lifetimes will also open up the possibility of studying these effects with other low frequency probes, ranging from measurements of magnetic susceptibility to scattering and transport methods. Recent theoretical[41] and experimental reports[42] suggest that the superconducting order parameter can be influenced also by the electromagnetic environment of an optical cavity. The observations made in in alkali doped fullerides[40] could be further expanded on by combining cavity settings with external driving, as a means to reduce the required excitation and hence dissipation and heating.

# **Acknowledgments**

The research leading to these results received funding from the European Research Council under the European Union's Seventh Framework Programme (FP7/2007-2013)/ERC Grant Agreement No. 319286 (QMAC). We acknowledge support from the Deutsche Forschungsgemeinschaft (DFG) via the Cluster of Excellence 'The Hamburg Centre for Ultrafast Imaging' (EXC 1074 – project ID 194651731). We thank Michael Volkmann for his technical assistance in the construction of the new optical apparatus presented in this work. We are also grateful to Elena König, Boris Fiedler and Birger Höhling for their support in the fabrication of the electronic transport samples, and to Jörg Harms for assistance with graphics.

# References (Main Text)

# Evidence for metastable photo-induced superconductivity in $K_3C_{60}$

M. Budden[1], T. Gebert[1], M. Buzzi[1], G. Jotzu[1], E. Wang[1], T. Matsuyama[1],

G. Meier[1], Y. Laplace[1], D. Pontiroli[2], M. Riccò[2], F. Schlawin[3], D. Jaksch[3], A. Cavalleri[1,3,*]

[1] Max Planck Institute for the Structure and Dynamics of Matter, 22761 Hamburg, Germany

[2] Dipartimento di Scienze Matematiche, Fisiche e Informatiche, Università degli Studi di Parma, Italy

[3] Department of Physics, Clarendon Laboratory, University of Oxford, Oxford OX1 3PU, United Kingdom

* e-mail: andrea.cavalleri@mpsd.mpg.de

## Supplementary Material

**S1 – Sample growth and characterization**

**S2 – Equilibrium optical properties and fitting models**

**S3 – Generation of picosecond mid-infrared pump pulses**

**S4 – Determination of the out-of-equilibrium optical response**

**S5 – Drude-Lorentz fits of the out-of-equilibrium optical response**

**S6 – Sample preparation for electrical transport measurements**

**S7 – Time-resolved electrical transport measurements**

**S8 – Pulse length dependence of the out-of-equilibrium metastable state**

**S9 – Relaxation dynamics of the out-of-equilibrium metastable state**

# S1 – Sample growth and characterization

The $K_3C_{60}$ powder pellets used in this work were prepared and characterized as previously reported in Refs. 1, 2. Finely ground $C_{60}$ powder and metallic potassium in stoichiometric amounts were placed in a vessel inside a Pyrex vial, evacuated to $10^{-6}$ mbar, and subsequently sealed. The two materials were heated at 523 K for 72 h and then at 623 K for 28 h. To ensure that $C_{60}$ was exposed only to a clean potassium vapor atmosphere, solid potassium and fullerene powder were kept separated during the heating cycle. The vial was then opened under inert atmosphere (in an Ar glove box with <0.1 ppm $O_2$ and $H_2O$) and the black powder was reground, pelletized and further annealed at 623 K for 5 days. This yielded phase pure $K_3C_{60}$ powders, as confirmed by powder X-ray diffraction measurements (Fig. S1a) that indicate an average grain size ranging from 100 nm to 400 nm. Figure S1b shows magnetic susceptibility measurements of the obtained $K_3C_{60}$ pellets upon cooling with an external magnetic field of zero (ZFC) and 400 A/m (FC). A critical temperature around 19.8 K can be extracted, which is in agreement with literature[3].

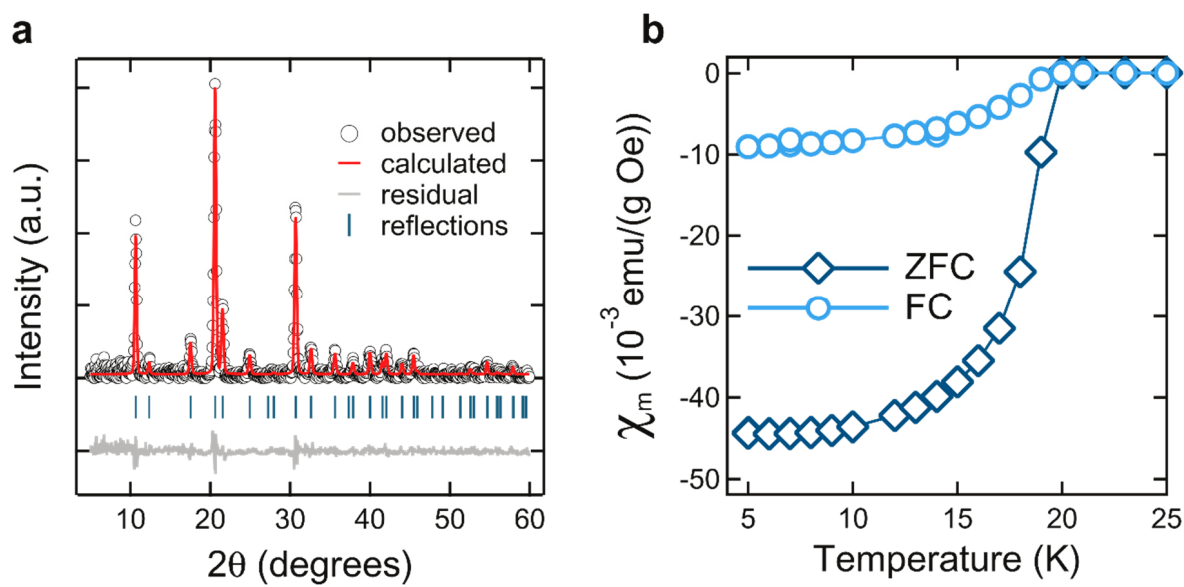

**Figure S1: a.** X-ray diffraction data and single f.c.c. phase Rietveld refinement for the $K_3C_{60}$ powder used in this work. **b.** Temperature dependence of the sample magnetic susceptibility measured by SQUID magnetometry upon cooling without (ZFC: zero field cooling) and with a magnetic field applied (FC: field cooling).

# S2 – Equilibrium optical properties and fitting models

The equilibrium optical properties of $K_3C_{60}$ were retrieved in a wide spectral range from 5 meV to 500 meV at temperatures between 25 K and 300 K. For this purpose, the reflectivity of the $K_3C_{60}$ pellet was measured by Fourier-transform infrared spectroscopy at the SISSI beamline (Elettra Synchrotron Facility, Trieste, Italy) using a commercial Fourier transform infrared spectrometer equipped with a microscope. The pellet was embedded in a sealed holder, pressed against a diamond window to obtain an optically flat interface, and attached to a helium cooled cryostat for temperature dependent measurements. To prevent degradation of the $K_3C_{60}$ pellets, all sample handling was performed in a glove box with Argon atmosphere (<0.1 ppm $O_2$ and $H_2O$).

The $K_3C_{60}$ reflectivity spectra were referenced against a gold mirror placed at the sample position. The low energy part of the spectrum (< 5 meV) was extrapolated using a Drude-Lorentz fitting while for the high energy side (> 500 meV) data on $K_3C_{60}$ single crystals was used[4,5]. The complex valued optical conductivity was retrieved through a Kramers-Kronig transformation for samples in contact with a transparent window[6].

Figure S2 shows the equilibrium optical properties of $K_3C_{60}$ at ambient pressure and at a fixed temperature T = 100 K. This and further data measured at different temperatures and pressures were already reported in Refs. 1, 2 and discussed also in comparison with data obtained from single crystals.

The measured conductivity spectra were fitted by a combination of a Drude term centered at ω = 0 denoting the free-carriers and a Lorentz oscillator reproducing the mid-infrared absorption at higher frequencies ($\omega_0 \sim$ 50 meV – 100 meV):

$$\sigma_1(\omega) + i\sigma_2(\omega) = \frac{\omega_p^2}{4\pi}\frac{1}{\gamma_D - i\omega} + \frac{\omega_{p,osc}^2}{4\pi}\frac{\omega}{i(\omega_{0,osc}^2 - \omega^2) + \gamma_{osc}\omega},$$

here $\omega_p$ and $\gamma_D$ are the plasma frequency and scattering rate of the Drude term, while $\omega_{p,osc}$, $\gamma_{osc}$, and $\omega_{0,osc}$ are the oscillator strength, the linewidth, and the resonance frequency of the additional Lorentz term. The result of the fit is shown as a solid blue curve in Figure S2. The equilibrium data reported here were used to normalize the transient optical spectra of $K_3C_{60}$ measured upon photoexcitation, as discussed in detail in section S4.

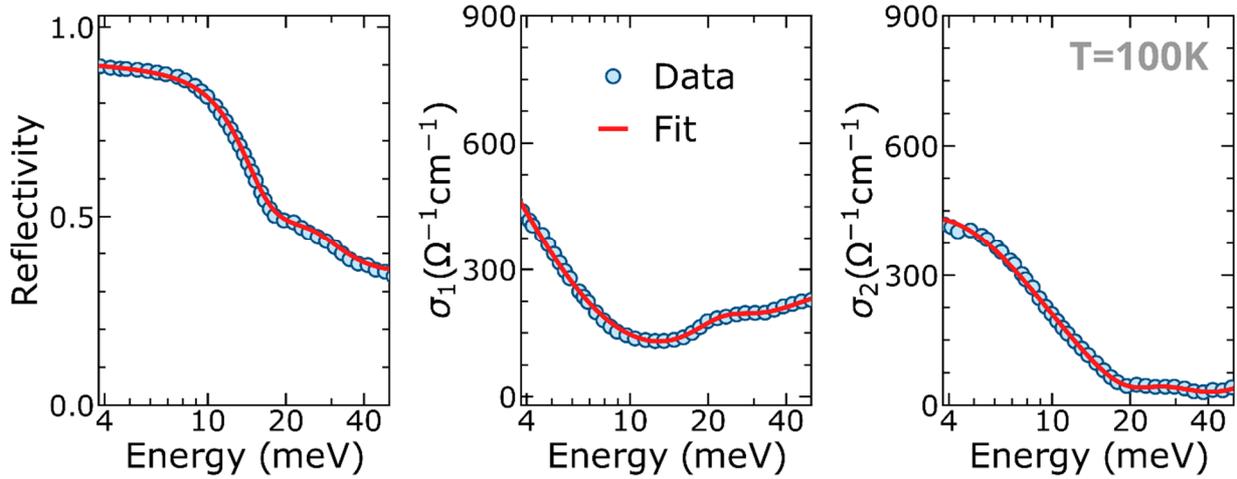

**Figure S2:** Equilibrium optical properties (reflectivity, real, and imaginary part of the optical conductivity) of $K_3C_{60}$ measured at a temperature of 100 K (blue dots). The red curve is a Drude-Lorentz fit to the optical conductivity as described in the text.

## S3 – Generation of picosecond mid-infrared pump pulses

The 1 ps long pump pulses centered at 7.3 µm wavelength used for the experiments reported in Figure 2 were obtained starting from ~100 fs long, 7.3 µm wavelength pulses generated by difference frequency mixing of the signal and idler outputs of a home built optical parametric amplifier (OPA) in a 0.5 mm thick GaSe crystal. The OPA was pumped with ~60 fs long pulses from a commercial Ti:$Al_2O_3$ regenerative amplifier (800 nm central wavelength). To obtain a pulse duration of ~1 ps, the 7.3 µm radiation was propagated through a highly dispersive 16 mm long $CaF_2$ rod. The spectrum of the pump pulses was characterized using a home built FTIR spectrometer. Their duration was measured by cross-correlation with a synchronized, 35 fs long, 800 nm wavelength pulse in a 50 µm thick GaSe crystal.

This experimental setup could not efficiently produce pulses with a duration longer than few picoseconds, and also lacked the flexibility of a continuously variable pulse duration. For this reason, the experiments reported in Figure 3, 4, and 5, were obtained with a new optical device based on $CO_2$ lasers that generated longer, narrow-band pulses centered at 10.6 µm wavelength, with fully tunable duration between 5 ps and 1.3 ns.

To perform mid-infrared pump, THz probe experiments, the pump pulses generated from the $CO_2$ laser system needed to be synchronized to the Ti:$Al_2O_3$ laser generating the THz probe light (see section S4). To achieve this, we developed the setup described in

Figure 3a. Here, ~150 fs long, 10.6 μm wavelength pulses were generated by difference frequency mixing of the signal and idler outputs of a home built OPA in a 1.5 mm thick GaSe crystal. The OPA was pumped with ~100 fs long pulses from a commercial Ti:Al$_2$O$_3$ regenerative amplifier (800 nm wavelength). A fraction of these femtosecond pulses (~60 pJ) were injected through the semitransparent front window (20 % transmission) into the cavity of a commercial transversely excited atmosphere (TEA) CO$_2$ laser. The injection of 10.6 μm wavelength seed pulses, induces a temporal mode locking resulting in a train of output pulses, which are synchronized to the femtosecond seed laser[7]. Because of the high finesse of the CO$_2$ laser oscillator cavity the seed pulses were spectrally filtered and the oscillator produced pulses with nanosecond duration. The most intense pulse from the train was selected with a custom designed CdTe Pockels cell and mid-infrared wire grid polarizers. The resulting output consisted of a single pulse with a duration of ~1.3 ns and a pulse energy of ~730 μJ per pulse at 18 Hz repetition rate. This pulse was then amplified further in a second ten-pass amplifier based on a modified commercial TEA CO$_2$ laser. The typical pulse energy achieved after the amplifier is ~11 mJ at 18 Hz repetition rate, with a pulse duration of ~1.3 ns.

The pulse duration of these 1.3 ns long pulses, was tuned using a combination of a plasma-mirror and -shutter, which allows "slicing" of the leading and trailing edge of the mid-infrared (MIR) pulses. For both plasma slicers we utilized semiconductor wafers transparent to the 10.6 μm radiation (Si, Ge, or CdTe) set at the Brewster's angle to suppress the reflection of the p-polarized MIR beam. A pair of time-delayed, intense femtosecond pulses (λ = 800 nm, 100 fs duration) was used to photoexcite the semiconductors to create an electron hole plasma at the surface that acts as an ultra-fast switchable mirror[8-10]. Varying the time delay between the two femtosecond pulses enabled us to tune their pulse duration between 5 ps and 1.3 ns. To compensate the path length in the cavities of the CO$_2$ laser and amplifier (~300 m), the slicing pulses were derived from a second Ti:Al$_2$O$_3$ amplifier. This was optically synchronized to the one used for generating the seed pulses for the CO$_2$ laser by seeding it with pulses from the same Ti:Al$_2$O$_3$ master oscillator.

The envelope of the sliced mid-infrared pulses is affected by the decay time of the electron/hole plasma in the semiconductor. When using n-doped silicon, the long carrier recombination time yielded pulses that have an almost flat-top shape for different pulse lengths up to ~300 ps. The pulse envelope of the generated pulses was measured by cross-correlation with a synchronized 100 fs long, 800 nm wavelength pulse in a 2 mm thick GaSe crystal (Figure S3).

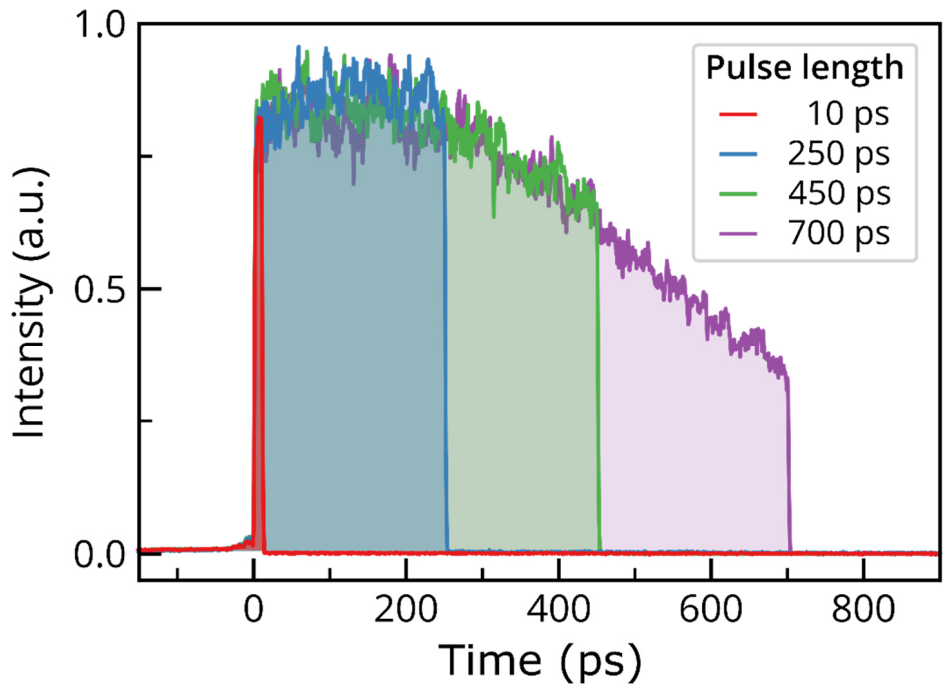

**Figure S3:** Time profile of the 10.6 μm pulses generated by slicing in two photo-irradiated n-doped silicon wafers. These traces are a result of a cross-correlation measurement following the procedure described in the text.

## S4 – Determination of the out-of-equilibrium optical response

The mid-infrared pump-THz probe experiments presented in Figures 1, 2, and 3 were performed on compacted $K_3C_{60}$ powder pellets pressed against a diamond window to ensure an optically flat interface. As $K_3C_{60}$ is water and oxygen sensitive, the pellets were sealed in an air-tight holder and all sample handling operations were performed in an Argon filled glove box with <0.1 ppm $O_2$ and $H_2O$. The sample holder was then installed at the end of a commercial Helium cold-finger (base temperature 5 K), to cool the pellets down to a temperature of 100 K.

The mid-infrared pump induced changes in the low frequency optical properties, were retrieved using transient THz time domain spectroscopy in two different experimental setups. The THz pulses were generated in a commercial photoconductive emitter or in a 0.2 mm thick (110)-cut GaP crystal using 800 nm pulses with a duration of 100 fs and 35 fs, respectively. In both cases, these 800 nm pulses were generated by the $Ti:Al_2O_3$ laser optically synchronized with the one producing the mid-infrared excitation pulses. The generated probe pulses were focused at normal incidence on the K$_3$C$_{60}$ sample, and their electric field profile was measured, after reflection from the sample, by electro-optic sampling in 1 mm thick ZnTe, or 0.2 mm thick GaP (110)-cut crystals. The ZnTe based setup had a measurement bandwidth ranging from 3.3 meV to 12 meV, while the GaP based one spanned the range between 4.1 meV to 29 meV. The time resolution of both setups is determined by the measurement bandwidth and is ~300 fs and ~150 fs, respectively.

To minimize the effects on the pump-probe time resolution due to the finite duration of the THz probe pulse, we performed the experiment as described in Refs. 11, 12. The transient reflected THz field at each time delay $\tau$ after excitation was obtained by fixing the delay $\tau$ between the pump pulse and the electro-optic sampling gate pulse, while scanning the delay $t$ of the single-cycle THz probe pulse.

The electric field reflected by the unperturbed sample, $E_R(t)$, and the pump-induced changes, $\Delta E_R(t,\tau)$ were simultaneously acquired at each time delay $\tau$ by acquiring the electro-optic sampling signals and chopping the pump and probe beams at different frequencies. Simultaneous measurement of the reference electric field $E_R(t)$ and the light-induced changes $\Delta E_R(t,\tau)$ avoids the introduction of possible phase artifacts (e.g. due to long term drifts) and is particularly useful when the measured electric field contains fast-varying frequencies. $E_R(t)$ and $\Delta E_R(t,\tau)$ were then independently Fourier transformed to obtain the complex-valued, frequency dependent $\tilde{E}_R(\omega)$ and $\Delta\tilde{E}_R(\omega,\tau)$. The photo-excited complex reflection coefficient $\tilde{r}(\omega,\tau)$ was determined by

$$\frac{\Delta\tilde{E}_R(\omega,\tau)}{\tilde{E}_R(\omega)} = \frac{\tilde{r}(\omega,\tau) - \tilde{r}_0(\omega)}{\tilde{r}_0(\omega)},$$

where $\tilde{r}_0(\omega)$ is the stationary reflection coefficient known from the equilibrium optical response (see Supplementary Section S2).

As the mid-infrared pump penetrated less ($d_{pump} \approx 0.2$ μm) than the THz probe ($d_{probe} \approx 0.6 - 0.9$ μm), these light-induced changes, measured at each pump-probe delay $\tau$, were reprocessed to take this mismatch into account. As the pump penetrates in the material, its intensity is reduced and it induces progressively weaker changes in the refractive index of the sample. A sketch of this scenario is shown in Figure S4, which was modelled by considering the probed depth of the material $d_{probe}$ as a stack of thin layers, with a homogeneous refractive index and assuming the excitation profile to follow an exponential decay. By calculating the complex reflection coefficient of this "multilayer" system with a characteristic matrix approach[13], the complex refractive index at the surface $\tilde{n}(\omega,\tau)$, can be self-consistently retrieved. From this, the complex conductivity for a homogeneously transformed volume was obtained as:

$$\sigma(\omega,\tau) = \frac{\omega}{4\pi i}[\tilde{n}(\omega,\tau)^2 - 1].$$

The only free parameter in this modelling is the intensity penetration depth of the mid-infrared pump, which is determined by the equilibrium intensity extinction coefficient at the pump wavelength, $\lambda_{pump}/4\pi Im(\tilde{n}_0(\omega=\omega_{pump}))$. The probe penetration depth $d_{probe}$ is a frequency- and time-dependent quantity that was self-consistently extracted from the transient response of the material $Im(\tilde{n}(\omega,\tau))$ through the multilayer modelling.

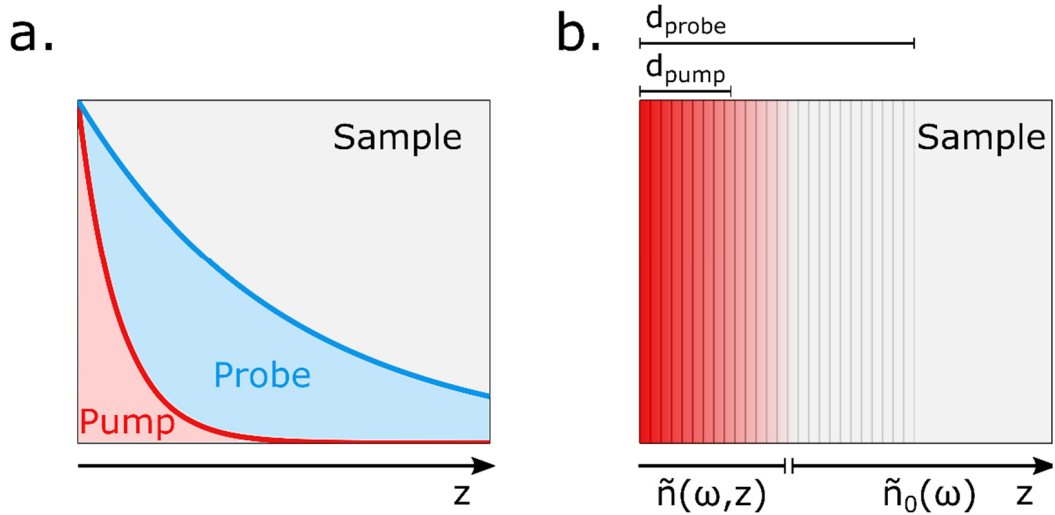

**Figure S4: a.** Schematics of pump-probe penetration depth mismatch. **b.** Multi-layer model with exponential decay used to calculate the pump-induced changes in the complex refractive index $\tilde{n}(\omega,\tau)$ for each pump-probe delay $\tau$. The transition from red to background (grey) represents the decaying pump-induced changes in $\tilde{n}(\omega,z)$.

# S5 – Drude-Lorentz fits of the out-of-equilibrium optical response

The out-of-equilibrium optical response of the photo-irradiated $K_3C_{60}$ pellets was modelled by fitting simultaneously the reflectivity and complex optical conductivity with the same Drude-Lorentz model used to fit the equilibrium response described in Section S2:

$$\sigma_1(\omega) + i\sigma_2(\omega) = \frac{\omega_p^2}{4\pi}\frac{1}{\gamma_D - i\omega} + \frac{\omega_{p,osc}^2}{4\pi}\frac{\omega}{i(\omega_{0,osc}^2 - \omega^2) + \gamma_{osc}\omega}$$

As previously reported[1], this model is also able to capture the photo-induced superconducting-like response of $K_3C_{60}$ as in the limit of $\gamma_D \to 0$ the Drude conductivity can capture the response of a superconductor below gap:

$$\sigma_1(\omega) + i\sigma_2(\omega) = \frac{\pi}{2}\frac{N_S e^2}{m}\delta[\omega = 0] + i\frac{N_S e^2}{m}\frac{1}{\omega} + \frac{\omega_{p,osc}^2}{4\pi}\frac{\omega}{i(\omega_{0,osc}^2 - \omega^2) + \gamma_{osc}\omega}$$

Here $N_S$, $e$, and $m$ are the superfluid density, electron charge, and mass, respectively. Furthermore, the transient nature of a photo-induced superconductor having a finite lifetime appears as a broadening of the zero-frequency Dirac delta[14].

Figure S5 shows representative fits to data measured at 100 K for three different time delays of -150, -100, and 10 ps. The fitting was performed keeping the parameters describing the mid-infrared polaronic band fixed, and letting only the Drude parameters $\omega_p$ and $\gamma_d$ free to vary. This fitting was performed on transient conductivity spectra for each time delay shown in Figure 4a. The obtained parameters were used to calculate the zero frequency extrapolated optical conductivity $\sigma_0$ and from this the "optical resistivity" $\rho_0$ as:

$$\rho_0 = 1/\sigma_0 = \lim_{\omega \to 0} 1/\sigma_1(\omega)$$

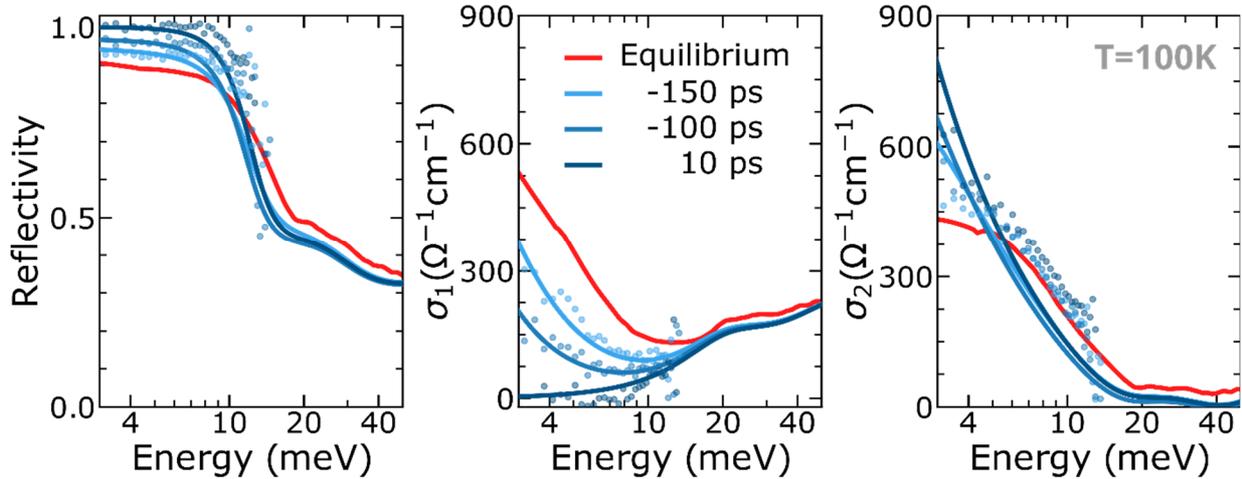

**Figure S5:** Measured optical properties (reflectivity, real, and imaginary part of the optical conductivity) of $K_3C_{60}$ measured at a temperature of 100 K at equilibrium (red curve) and at three different time delays of -150, -100, 10 ps (blue filled symbols). These data were acquired with a pump-pulse duration of 200 ps and by definition the time delay is zero when the pulse ends (cf. Figure 3c). The solid lines are Drude-Lorentz fits to the transient optical properties.

## S6 – Sample preparation for electrical transport measurements

For time-resolved electrical transport measurements, pellets of $K_3C_{60}$ were integrated into a sample carrier with patterned microstrip transmission lines. Fig. S6 shows a picture of the sample carrier and a sketch of its cross-section. The four Ti (10 nm)/Au (270 nm) microstrip structures were grown using a combination of e-beam evaporation, laser-lithography, and lift-off processing, on a 500 μm thick diamond substrate, transparent to the 10.6 μm radiation. The wave impedance of the transmission lines was adapted to 50 Ω. A pellet of 1 mm diameter and ~75 μm thickness was made from $K_3C_{60}$ powders with a pellet die and a manual press. All the sample handling operations were carried out in an Argon filled glove box with <0.1 ppm $O_2$ and $H_2O$ to prevent sample oxidation. To ensure good electrical contact between the polycrystalline pellet and the transmission lines at low temperatures, a layer of 2 μm thick Indium was deposited on the inner parts of the Au transmission lines by an additional lithography and lift-off step. To reproducibly position the pellet in the center of the four microstrips, a ~50 μm thick layer of photoresist (SU8) with a 1 mm central bore was deposited. The $K_3C_{60}$ pellet enclosed in the photoresist layer, was then capped with a 500 μm thick sapphire plate and sealed with vacuum compatible glue. This assembly was placed on a copper holder and then installed at the end of a commercial LHe cold-finger, in order to cool the $K_3C_{60}$ pellets down to a base temperature of 5 K. The microstrips were terminated to SMP connectors to connect them to 50 Ω wave impedance coaxial cables that are thermally anchored on the cold finger, and routed to the outside of the cryostat.

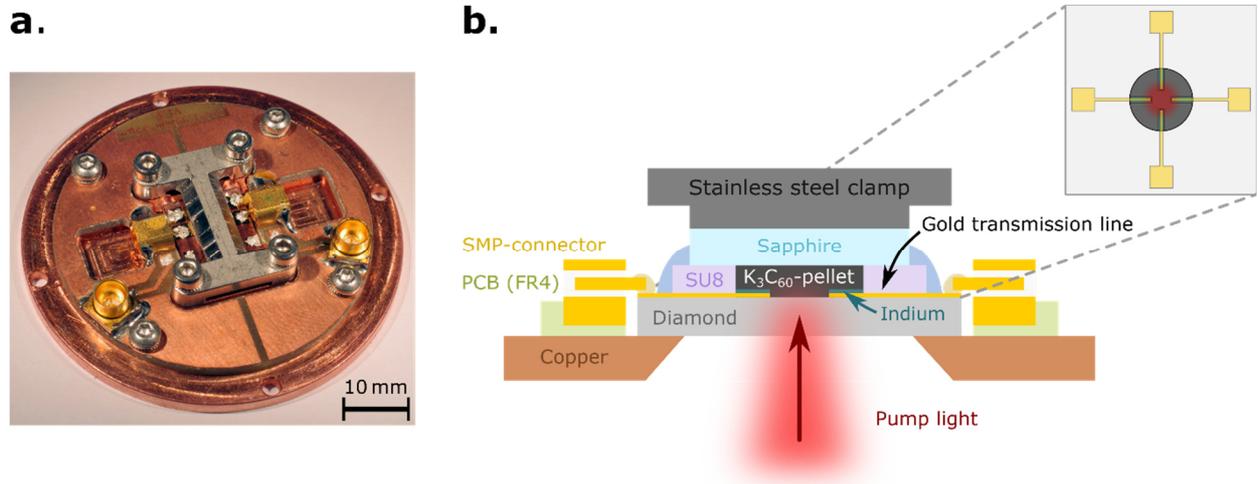

**Figure S6: a.** Photograph of the sample carrier used for transport measurements. **b.** Schematic section-view of the $K_3C_{60}$ sample assembly (not to scale). The pulsed mid-infrared excitation reaches the sample from below. The upper right panel shows a sketch of the electrode geometry similar to the one used in both four- and two-terminal measurements.

## S7 – Time-resolved electrical transport measurements

Standard four-terminal resistance measurements (Figure S7a) were performed on $K_3C_{60}$ pellets encapsulated in the high-frequency sample carrier using a lock-in technique with a sinusoidal excitation current of amplitude $I_{source} = 1$ µA at a frequency of 300 Hz. While this technique provides accurate values of the low frequency sample resistance $R_{sample}$ it could only be applied up to frequencies of a few MHz due to the limited bandwidth of the high input impedance differential amplifiers as well as parasitic capacitances in the circuit.

The lifetime of the light-induced superconducting state is several nanoseconds, hence four-terminal measurements are too slow to probe transport properties on these time scales. To overcome this issue, we performed high-frequency two-terminal resistance measurements. These measurements were conducted by launching a 1 ns long voltage pulse from a commercial pulse generator through the microstrip transmission lines and the $K_3C_{60}$ pellet in the sample carrier. A gated integrator was then used to detect the amplitudes of the injected voltage pulse $V_{in}(t)$, and of the one transmitted through the sample $V_{out}(t)$. An equivalent circuit diagram of this two-terminal measurement is shown in Figure S7b. These measurements allow to retrieve the resistance $R^*_{2P} = R_{sample} + R_{C1} + R_{C2}$ which includes contributions from wiring and contact resistances. As the injected voltage pulse propagates, it can be reflected at the input and output terminals of $R^*_{2P}$ due to a possible impedance mismatch. To account for this, it is convenient to describe the propagation of the pulse through the network using the two-port scattering matrix formalism[15], from which the total resistance of the sample $R^*_{2P}$ can be retrieved as:

$$R^*_{2P} = R_{sample} + R_{C1} + R_{C2} = 2\left(\frac{V_{in}}{V_{out}} - 1\right) R_m.$$

Here, $V_{in}$ and $V_{out}$ are the amplitudes of the injected pulse and of the pulse transmitted through the sample while $R_m$ = 50 Ω is the input impedance of the gated integrator used for signal detection. This result was quantitatively verified by simulating the equivalent circuit with the software QucsStudio starting from measured values of the elements in the circuits (not shown).

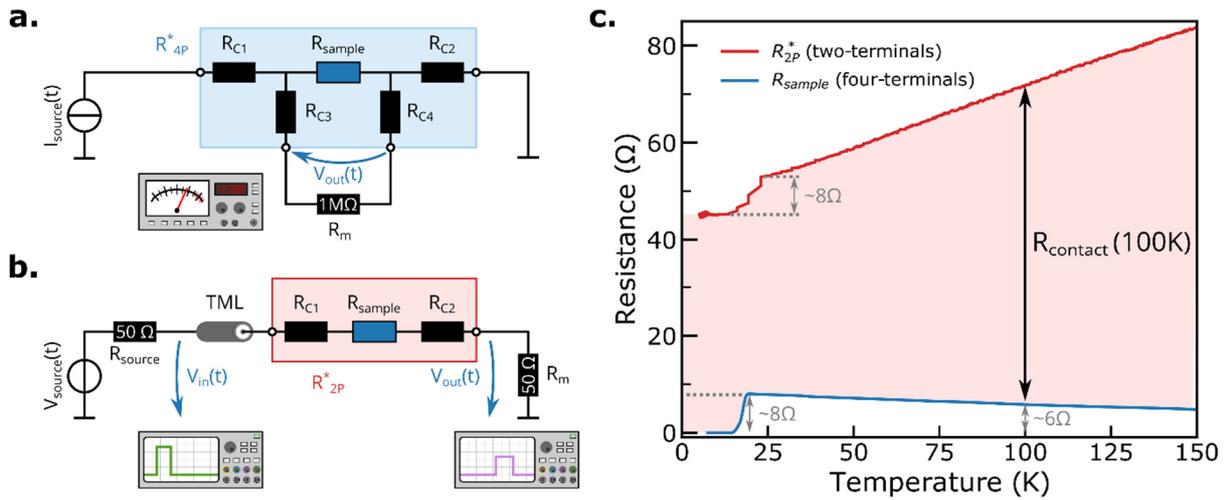

**Figure S7: a.** Equivalent circuit of a four-terminal resistance measurement. A constant amplitude sinusoidal current is injected into $R_{sample}$ through the contact resistances $R_{C1}$, and $R_{C2}$. The voltage drop across $R_{sample}$ is detected by a high input impedance lock-in amplifier through the contact resistances $R_{C3}$ and $R_{C4}$. The sample resistance can be directly retrieved as $V_{out}/I_{source}$. **b.** Equivalent circuit for two-terminal measurements. The sample with intrinsic resistance $R_{sample}$ and contact resistances $R_{C1}$ and $R_{C2}$ is connected via two terminals. TML indicates the coaxial cable which acts as a transmission line for high frequency signals. **c.** Temperature dependent resistance measured in a two- and four-terminal measurement. The temperature-dependent contact resistance $R_{contact} = R_{C1} + R_{C2}$ is displayed as red shading inbetween the graphs. Both measurements were performed on the same type of pellet with identical preparation procedure and equal electrode geometry.

Extracting $R_{sample}$ from $R_{2P}^*$ requires knowledge of the contact resistance from the low-frequency four-terminal measurements. In order to calibrate the contact resistance $R_C = R_{C1} + R_{C2}$ we compared the temperature dependence of the resistance $R_{2P}^*$ (measured in the pulsed two-terminal geometry, red curve in Fig. S7c) to that of the sample resistance $R_{sample}$ (measured in the four-terminal geometry, blue curve in Fig. S1c). In both measurements, the equilibrium superconducting transition of K$_3$C$_{60}$ is observed as a drop in the measured resistance of ~8 Ω as the sample becomes superconducting. Hence, the two- and four-terminal measurements are equally sensitive to changes in $R_{sample}$ and are only offset by the contact resistance $R_C$ (red shaded area in Figure S1c), allowing us to extract $R_{sample}$ from the high-frequency two-terminal measurements.

The time-resolved resistance measurements of the photo-irradiated K$_3$C$_{60}$ pellet shown in Figure 4b were performed at 100 K by repeating the pulsed two-terminal measurements at different time delays after photoexcitation. This was achieved by electronically synchronizing the pulse generator that provided the probe voltage pulse to the laser system. These measurements yielded the total resistance $R_{2P}^*$, from which the contact resistance $R_C(100\ K) = 66\ \Omega$ was subtracted to obtain the time-dependent resistance of the K$_3$C$_{60}$ sample alone. Note that the change of the resistance in the time-resolved experiment at T = 100 K is ΔR = 6 Ω (Figure 4), which is what is quantitatively expected from the DC four-terminal measurement if the sample turns superconducting at this temperature.

## S8 – Pulse length dependence of the out-of-equilibrium metastable state

In analogy to Figure 5, Figure S8 shows excitation-pulse length dependent data extracted from transient THz time domain spectroscopy measurements. Here, we plot the reduction of spectral weight in the real part of the optical conductivity (lower panels) as

$$\Delta\sigma_1 = \int_{2.2 meV/\hbar}^{10 meV/\hbar} \sigma_1^{trans}(\omega) - \sigma_1^{eq}(\omega)\ d\omega,$$

upon excitation with mid-infrared pulses of different duration varying between 250 ps and 700 ps (top panels). The excitation fluence was kept constant to a value of 22 mJ/cm². The data indicates that $\Delta\sigma_1$ measured 100 ps and 500 ps after excitation is independent of the excitation pulse duration. This is in agreement with the pulse-length dependent data presented in Fig. 5 and measured by electronic transport. Pump pulse lengths shorter than 250 ps were not measured here since the fluence of 22 mJ/cm² could not be achieved for these shorter pulses.

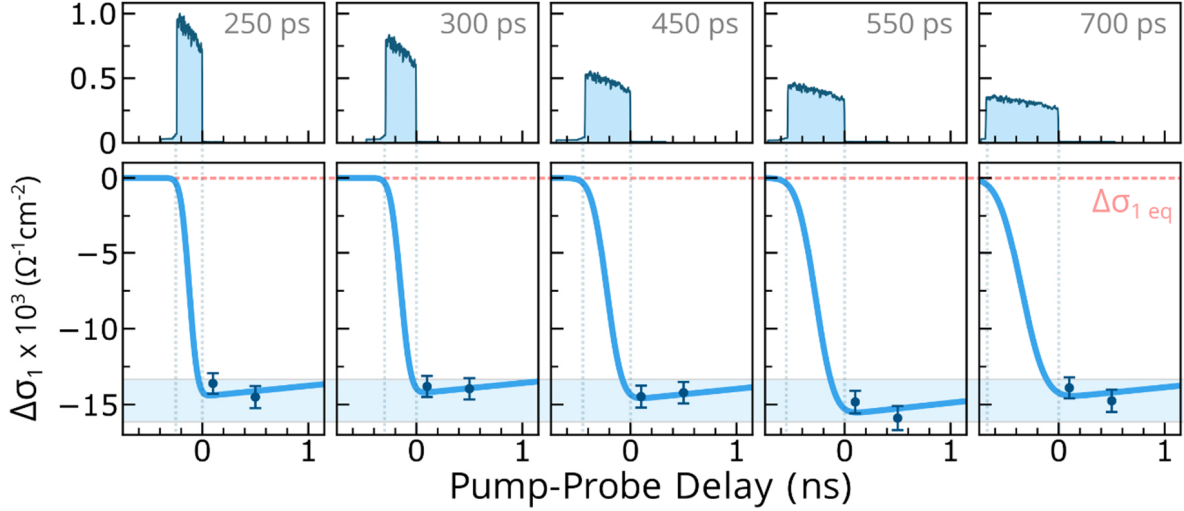

**Figure S8: (lower panels)** Pump pulse length dependence of the spectral weight loss $\Delta\sigma_1$ (as defined in the text) induced upon photoexcitation and measured at time delays of 100 ps and 500 ps for different pump pulse durations. **(upper panels)** Excitation pump pulse profiles of the mid infrared pulses measured via cross-correlation (see Section S3). The blue curves are a guide to the eye indicating the typical time dependence of the signal. All measurements were performed at a constant fluence of 22 mJ/cm², excitation wavelength of 10.6 µm, and base temperature T = 100 K.

## S9 – Relaxation dynamics of the out-of-equilibrium metastable state

The relaxation dynamics of the metastable photo-induced superconducting state in $K_3C_{60}$ were measured both optically and electronically. Figure S9a displays the time evolution of the photo-induced reduction of spectral weight in the real part of the optical conductivity $\Delta\sigma_1$ in the region between 2.2 meV and 10 meV. These measurements were performed on two different $K_3C_{60}$ samples with a 10.6 µm pump pulse of 200 ps duration and at a fluence of 30 mJ/cm². The relaxation behavior can be modelled with an exponential decay with a time constant of ~12 ns (dashed line). To achieve optical delays longer than 2 ns we introduced additional fixed retardation lines into our setup to extend the range of the continuously tunable optical delay stages.

Figure S9b displays the time-dependent resistance of the laser irradiated $K_3C_{60}$ pellet obtained from transient two-terminal measurements. The graph shows the same data presented in Fig. 4b. The inset, displays the time-dependence of the photo-induced resistance up to a time delay of 1 μs. The relaxation can be modelled with a double exponential decay (light blue solid line) one with a faster time constant $\tau_1 \sim 30$ ns (compatible with what is shown in Figure S9a) and a second slower one with $\tau_2 \sim 550$ ns.

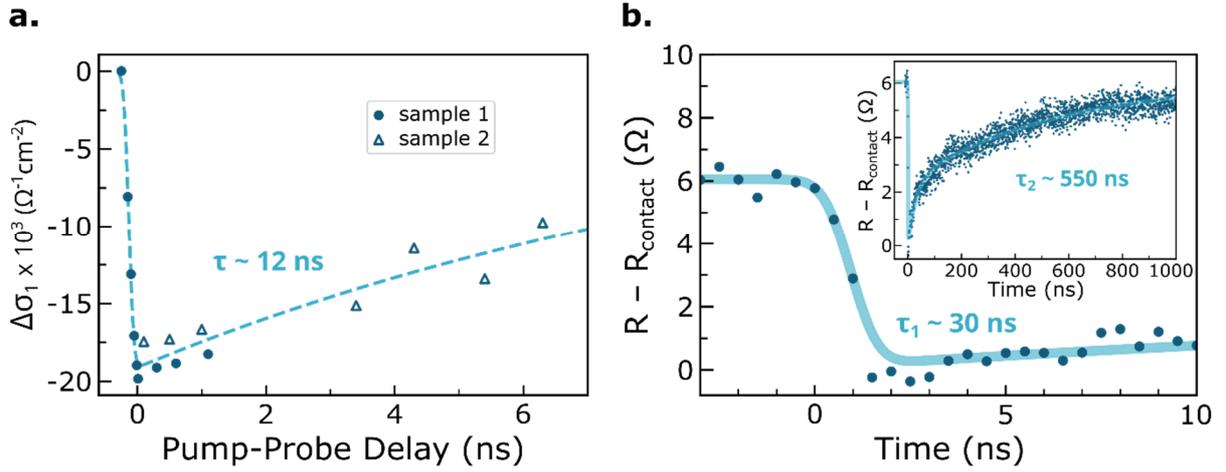

**Figure S9: a.** Time dependent spectral weight loss $\Delta\sigma_1$ as defined in supplementary section S8, measured on two different $K_3C_{60}$ samples. The dashed line is a fit with a single exponential decay yielding a time constant $\tau_1 \sim 12$ ns. **b.** Time-dependent resistance of a laser irradiated $K_3C_{60}$ pellet obtained from high-frequency two-terminal transport measurements. The graph shows the same data as in Figure 4b. The inset displays the relaxation dynamics of the signal on a longer timescale. The light blue solid line is a fit with a double exponential decay function ($\tau_1 \sim 30$ ns and $\tau_2 \sim 550$ ns). All measurements were performed at an excitation wavelength of 10.6 μm, and base temperature T=100 K.

# References (Supplementary Material)